\documentclass{aa}
\usepackage{graphics}
\usepackage{rotating}
\begin{document}
\newcommand{\mcol}[3]{\multicolumn{#1}{#2}{#3} }
\newcommand{\struut}{\rule[-2ex]{0ex}{5.2ex}}
\newcommand{\struutup}{\rule{0ex}{3.2ex}}
\newcommand{\struutdown}{\rule[-2ex]{0ex}{2ex}}

   \thesaurus{06  
              (03.20.4 ; 05.01.1;  
               08.02.6 ; 08.06.3; 03.19.2 )} 
%
   \title{CCD Photometry and Astrometry for Visual Double and Multiple 
   Stars of the HIPPARCOS Catalogue\thanks{Based on observations made at La Silla (ESO, Chile 
   - Key Programme 7-009-49K), Observatoire de Haute-Provence (OHP), Calar Alto (CLA), La
   Palma (LPL) and Jungfraujoch (JFJ) Observatories }}
   \subtitle{I. Presentation of the large scale project}

   \author{Oblak, E. \inst{1}, Lampens, P. \inst{2}, Cuypers, J. \inst{2},
           Halbwachs, J.L. \inst{3}, Mart\'{\i}n, E. \inst{4,5}, Seggewiss, W. \inst{6}, 
	   Sinachopoulos, D. \inst{2}, van Dessel, E.L. \inst{2}, Chareton, M. \inst{1}
          \and 
           Duval, D. \inst{2}
          }

   \offprints{E. Oblak}

   \institute{
             Observatoire de Besan\c{c}on, 
             41 bis, avenue de l'Observatoire, BP 1615, F-25010 Besan\c{c}on Cedex, France\\
             email: oblak@obs-besancon.fr
         \and
             Koninklijke Sterrenwacht van Belgi\"e,
             Ringlaan 3, B-1180 Brussel, Belgium\\
             email: patricia.lampens@oma.be
         \and
             Observatoire de Strasbourg, 
             11, rue de l'Universit\'e, F-67000 Strasbourg, France\\
             email: halbwachs@astro.u-strasbg.fr
         \and
             University of California at Berkeley, 
	     601 Campbell Hall, Berkeley CA 94720, USA\\
             email: ege@popsicle.berkeley.edu
         \and
             Instituto de Astrofisica de Canarias, 
	     38200 La Laguna, Tenerife, Spain
         \and
             Observatorium Hoher List der Universit\"ats-Sternwarte Bonn, 
             D-54550 Daun, Germany\\
             email: seggewis@astro.uni-bonn.de
            }

   \authorrunning{Oblak, E., Lampens, P. et al.}
   \titlerunning{CCD results for HIPPARCOS visual double and multiple stars}
   \date{Received date/ Accepted date}

   \maketitle

   \begin{abstract}

  A description is given of the activities of an international working
  group created with the aim of obtaining both photometric and astrometric
  observations of visual double and multiple stars with angular separations 
  in the range of one to fifteen arcseconds, that formed part of the
  HIPPARCOS Input Catalogue. The scientific aims and realisations 
  of this European network are given. About 50~observational missions have been
  carried out in both hemispheres according to a pre-defined protocol. 
  We describe the general and specifically designed methods used for the reduction of 
  large amounts of CCD observations of double stars and give an outline of the 
  results already presented and soon to be expected.

   \end{abstract}


\section{Introduction}\label{sec:intro}


\subsection{A new project}\label{ssec:project}

   Mid-1990 a large-scale project was started for combining the efforts of scientists in six 
   European countries, from ten laboratories, with the goal to obtain accurate 
   ground-based photometric and astrometric information on visual double and multiple stars: the
   European Network of Laboratories "Visual Double Stars" was founded. Both hemispheres
   would be covered (Oblak et al. \cite{obl192}). In 1992 a key programme was introduced at the 
   European Southern Observatory (ESO) aiming at obtaining the photometry and the astrometry of 
   visual double stars in the southern hemisphere: photoelectric and CCD observations would be 
   obtained to complement the HIPPARCOS space observations on such systems (Oblak et al. 
   \cite{obl392}). This paper is meant to be a general introduction to this vast observational effort
   : its aim is to extensively report on the scientific goals (Sect.~\ref{sec:intro}) 
   and the technical aspects (Sects.~\ref{sec:obs}, ~\ref{sec:met}, ~\ref{sec:red}) as well as to 
   introduce a series of forthcoming data papers (Sect.~\ref{sec:con}).
   Our programme is defined in Sect.~\ref{sec:obs}. A large part consists
   in the description of the observational protocol (Sect.~\ref{sec:met}) and the introduction of
   an original reduction method specifically developed for this programme (Sect.~\ref{sec:red}), 
   some of the more important aspects to consider in order to finally obtain data of a quality 
   that eventually can sustain the comparison with space projects (such as HIPPARCOS). General 
   conclusions and future prospects are formulated at the end.
 
\subsection{The scientific goals}\label{ssec:goals}
   It is common knowledge that (visual) binary and multiple stars are prime targets for determining
   and calibrating basic stellar physics in general. In the first place, they serve to determine the masses and
   to calibrate the mass-luminosity relation.
   This is possible with sufficient accuracy, say better than 10\%, under optimum conditions only, i.e. for
   a visual binary that is both sufficiently nearby and orbiting with a short period. For example, making use 
   of the new Hipparcos absolute parallaxes, only 55 of the more than 1000 previously known orbital pairs 
   satisfy the condition 
   of accuracy on each component mass better than 15\% (Lampens et al. \cite{lam297}). Since angular separations
   are generally below 1\arcsec, these are the "close" visual binaries. The photometry on the individual components
   for these systems almost completely relies on visual estimates of their magnitude differences.
   In the second place, visual binaries also serve to calibrate other stellar parameters since their common formation
   implies a common origin (same overall metallicity) and a same age (sharing a common isochrone), conditions
   that may even apply stricter than for open cluster members. Elimination of those generally badly known 
   parameters allows to focus on the
   remaining ones that can therefore be investigated in an independent way. Such testcases may also be found
   among wider or "intermediate" visual binaries that have longer periods, separations larger than 1\arcsec, with
   high confidence that their components could not have influenced each other's evolution.
   For these systems however, the physical association should be clearly established (i.e. this should not concern too 
   wide pairs). The existing photometry of the components with separations less than 10\arcsec~still relies 
   mostly on visual estimates of $\Delta m$ but also on the area-scanning technique (Franz \cite{fra66}, Rakos et al. 
   \cite{rak82}) for some hundreds of visual double stars. The conventional photoelectric technique cannot be trusted 
   in this range of separations.\\
   In general, the vast majority of stars are members of binary or multiple systems. Recent surveys of high 
   astrometric quality, both from space (Lindegren \cite{lin97}) or from the ground, show clear
   evidence that improving the resolution of the instruments generates an increasing number of new detections and
   that the frequency of binaries is probably underestimated.
   The correct determination of the frequency as a function of stellar parameters such as spectral type, luminosity,
   population is a major constraint for the modeling of, for example, the galatic content and structure.
   Although more poorly known than their "single" counterparts - partly because 
   the research is easily biased by the employed techniques and by the complication of the observational data's 
   interpretation due to the presence of the companion stars -  
   the binary stars deserve to be studied "in their own right".\\ 
   It is therefore still a crucial matter to investigate the fundamental properties and 
   the typical characteristics of double and multiple stars of all classes. The determination of the distribution functions
   of a minimal set of basic parameters such as true separations (linked with angular separations), mass and 
   luminosity ratios (linked with differences
   of magnitudes) and differences of temperatures (linked with differences of colour indices) defines 
   the context of this observational work.\\
   
 
   The aim of our programme is, more specifically, to acquire and analyse the astrometric and astrophysical 
   information of the individual components of visual binaries measured by the
   HIPPARCOS satellite mission. Regarding the astrometry, confrontation of recent versus past 
   astrometric observations may lead to detect or to contradict orbital motions (e.g.
   Brosche et al. \cite{br92}, Bauer et al. \cite{bau94}). This is especially important for the
   intermediate and wide pairs since a study of their relative proper motions is the only way to discriminate
   between physical and orbital pairs. Regarding the photometry, we believe that 
   complementary accurate photometric multi-colour data with reliable astrophysical content 
   are needed for a well-chosen sample of binaries and multiple systems for which good 
   quality astrometric data already exist. Indeed, although detection of new 
   double and multiple stars from astrometric programmes is 
   continuing at a high rate due to their improving resolution, this high quality astrometry 
   is generally coupled to a scarce photometry or at best a poor quality photometry of the 
   {\it individual components} compared to that of the joint photometry on the system. 
   The reason is that accurate brightnesses and colours of the components of visual 
   systems can be obtained by conventional photoelectric aperture photometry in good 
   conditions only if the separation is larger than the size of the used diaphragms 
   (typically 11-13\arcsec) and if sufficient care is also taken when measuring the 
   sky contribution. At closer separations the photometric information on the components 
   is often either inaccurate or lacking: global photometry may exist but it has to be 
   combined with visual estimates of the differential magnitudes (such estimates are poorly
   known since they can be as much as 0.5 mag off) to obtain component magnitudes. At even closer 
   separations and at the decimag accuracy level, neither are speckle differential magnitudes 
   easily obtainable (Carbillet et al. \cite{car96}; Ten~Brummelaar et al. \cite{ten96}).
   The ($B_T$, $V_T$) photometry of the vast majority of the stars in the
   HIPPARCOS catalogue was obtained in the course of the TYCHO programme
  (ESA \cite{esa97}), at least for systems with separations wider than 3\arcsec. 
   Unfortunately, the Tycho photometry is not very accurate
   (the median precision is only 0.10 mag in $B_T - V_T$), and, moreover, the
   magnitude measurements of the double star components are contaminated by the
   companions (Halbwachs et al. \cite{halb97}).\\ 
   
   More generally, photographic magnitudes are found in very large double star catalogues 
   such as the Washington Double Star Catalogue at USNO (Worley \& Douglass \cite{wor97}) and the 
   Catalogue of Components of Double and Multiple Stars at ROB (CCDM, Dommanget \& Nys \cite{dom95}). 
   The lack of accurate photometric data is reflected by 
   the simple fact that much less than $10\%$ of the systems listed in the CCDM (Dommanget \& Nys \cite{dom95}) 
   and $\approx~10\%$ of the systems catalogued in the Annex of Double and Multiple Stars of the Hipparcos 
   Input Catalogue (Turon et al. \cite{tur92}) have photoelectric photometry for both components 
   (see the 'Catalogue Photom\'etrique des Syst\`emes Doubles et Multiples', CPSDM, 
   Oblak \cite{obl88}).\\
   Nowadays, observations made with CCD detectors permit to obtain accurate individual 
   photometric data in a separation range where previous techniques failed
   (Sinachopoulos \& Seggewiss \cite{sin89}; Argue et al. \cite{arg92}; 
   Nakos et al. \cite{nak97}). We applied this observational technique to obtain
   the relevant data for each component of "intermediate" visual pairs
   (defined as having angular separations between 1\arcsec and 15\arcsec) in parallel with 
   the conventional photoelectric technique for the "wide" visual pairs (with angular 
   separations larger than 12\arcsec).


\section{The observational programme}\label{sec:obs}

\subsection{General description}\label{ssec:desc}
   
 A large programme for the systematic acquisition of accurate, homogeneous
 photometric colour indices of the components of several thousands of double and 
 multiple systems was thus set up in both hemispheres with the following principal aims:\\
 
  - to construct a basic sample of nearby double stars
    with complete astrometric and photometric information
    for each component of the system. By choosing those
    systems that belong to the HIPPARCOS programme, we made sure that 
    the full astrometric information would be measured in space. 
    The HIPPARCOS proper motions and parallaxes 
    may now help to define "clean" samples (by filtering out systems that are
    most probably optical (Dommanget \cite{dom55}, \cite{dom56}, Brosche et al.
    \cite{br92}) as a function of 
    parallax (i.e. distance-limited samples). The new photometric data 
    will supplement the Hipparcos magnitudes with astrophysically significant 
    colours that, once calibrated, will provide us hopefully with additional
    information such as temperature, gravity or metallicity (Oblak \& Lampens \cite{obl292}).\\
    
  - to improve the accuracy of the component photometric data for a
    large sample of visual double stars in view of applications that
    concern the distributions of true separations, mass ratios and colour differences.
    The usefulness of accurate component photometry is furthermore also
    evident in several other previously described applications, e.g.
    luminosity calibrations, age and evolution determinations, etc.\\
  
  In addition, we also provide accurate astrometric and
  photometric data for components that, for one reason or another, were
  not succesfully measured or were "missed" by HIPPARCOS. This may refer to 
  components with angular separations larger than 10\arcsec~  
  not included in the Input Catalogue, to components with angular separations 
  comparable to or larger than the half-width of the 'instantaneous field of view' (IFOV)
  ($\geq$ 10\arcsec, i.e. a two-pointing double) for which the resulting astrometry/photometry 
  may be perturbed and to components fainter than the companion star by more 
  than approx. 3.5 mag or to components of those systems that were too difficult 
  to treat and without solution.

\begin{table}[h]
\caption{{\bf Time allocation for CCD and conventional photometry: South 
(Key Programme included)}} 
\vspace{5mm}
\begin{tabular}{|c|cccc| }
\hline 
\mcol{1}{|c|}{Type}                            
& \mcol{1}{c}{Date}
& \mcol{1}{c}{Observer}
& \mcol{1}{c}{Per}
& \mcol{1}{c|}{N.N.\struut} \\
\hline 
{\bf CCD} & DUT 0.9m &&& \struut \\
\hline
& Oct 91 & J. CUYPERS  & 48 & 10 \struutup \\
& Feb 92 & W. SEGGEWISS  & 48 & 7 \\
& May 92 & J. CUYPERS  & 49 & 5\\
& Aug 92 & P. LAMPENS & 49 & 5\\
& Nov 92 & E. OBLAK & 50 & 5\\
& Mar 93 & J. CUYPERS  & 50 & 5\\
& Jun 93 & SINACHOPOULOS  & 51 & 4\\
& Aug 93 & M. BURGER & 51 & 6\\
& Dec 93 & P. LAMPENS & 52 & 5\\
& Feb 94 & SINACHOPOULOS & 52 & 5\\
& May 94 & J.L. HALBWACHS  & 53 & 5\\
& Aug 94 & P. LAMPENS & 53 & 5\\
& Nov 94 & E. OBLAK & 54 & 5\\
& Jan 95 & E. OBLAK & 54 & \struutdown 6\\
\hline
{\bf CVT} & ESO 0.5/1m &&& \struut \\
\hline
& May 92 & J. CUYPERS  & 49 \struutup & 2\\
& Aug 92 & P. LAMPENS  & 49 &  3\\
& Nov 92 & E. OBLAK & 50 & 5\\
& Mar 93 & J. CUYPERS & 50 & 4 \\
& Jun 93 & SINACHOPOULOS & 51 & 3\\
& Aug 93 & M. BURGER  & 51 & 4\\
& Dec 93 & P. LAMPENS & 52 & 4\\
& Feb 94 & SINACHOPOULOS & 52 & 3\\
& Aug 94 & P. LAMPENS & 53 & 5 \\
& Nov 94 & E. OBLAK & 54 & \struutdown 4\\
\hline
\end{tabular} 
\end{table}

\subsection{Selection of programme stars}\label{ssec:sel}
 
 The selection of the programme was made starting from 11434 double systems, 1960 triple 
 systems, 536 quadruple and 237 multiple systems of the Annex of Double and Multiple Stars,
 containing a majority of objects within a distance of 500 pc (Turon et al. \cite{tur92}). 
 Systems for which the component photometric information was lacking or poor have been selected 
 by cross-examination with the CPSDM (Oblak \cite{obl88}).
 This catalogue contains all information in three photometric systems (UBV, Geneva, Str\"omgren)
 for visual double and multiple systems, with indications on which components have been
 observed. We eliminated a small number of systems for which all the known components already
 have complete and precise photometric measurements: it is the case for $\approx$ 11\% out of the
 11853 systems listed in this catalogue. This concerns the measurements of 237 systems in the Geneva 
 photometric system, 948 systems in the Str\"omgren photometric system and 1540 systems in the UBV system
 (Oblak and Mermilliod, \cite{om88}; Oblak et al., \cite{obl393}). The gross of the data regarding these 
 wide visual double stars (separations larger than 10-12\arcsec) comes from works such as done by Lindroos 
 (\cite{lin81}, \cite{lin83}, \cite{lin85}), Olsen (\cite{ol82a}, \cite{ol82b}), Sinachopoulos 
 (\cite{si89}, \cite{sin90}) and Wallenquist (\cite{wal81}). \\ 
 We selected all systems with angular separations $> 1$\arcsec~ for which either not all 
 components had been observed or whose differential magnitudes were insufficiently precise 
 for extraction of astrophysical quantities. We found that differential colour indices are 
 almost nonexistent in the separation range 1\arcsec~- 12\arcsec. Some 10\%  of visual double stars,
 generally the ones with separations larger than 10-12\arcsec, have colour indices for both components.
 Our programme consisted of northern ($\delta_{A} > -10 \degr$) and southern 
 samples ($\delta_{A} \leq +10 \degr$) to be measured in various photometric campaigns 
 and in both hemispheres. The overlapping zone in declination was observed once 
 according to feasibility.\\
 Since both conventional (CVT) and CCD photometry were used, the samples were also split
 with respect to angular separation, with a common intersection between 12\arcsec~ and 
 15\arcsec~ for calibration purposes. 
 Systems on the CCD observational programme had to satisfy the following criteria:  
 \begin{list} {}{\setlength{\parskip}{0mm} \setlength{\parsep}{0mm}
 \setlength{\itemsep}{0mm}  \setlength{\topsep}{0mm}}
 \item $ 1\arcsec~ < {\rm separation} \leq 15\arcsec$,  \item $ 0 \leq \Delta {\rm m} < 3$ mag, 
 \item lacking component photometry.  \end{list} 
 Our goal was thus to obtain accurate magnitudes and colours for the components 
 of some 3000 HIPPARCOS double stars and some 600 multiple stars using
 both techniques (Oblak \& Lampens \cite{obl292}).\\  
   
\begin{table}[t]
\caption{{\bf Time allocation for CCD and conventional photometry: North}} 
\vspace{5mm}

\begin{tabular}{|c|cccc| }
\hline 
\mcol{1}{|c|}{{\bf Type}}                            
& \mcol{1}{c}{{\bf Date}/Filters}
& \mcol{1}{c}{{\bf Observer}}
& \mcol{1}{c}{{\bf Site}}
& \mcol{1}{c|}{{\bf N.N.} \struut} \\
\hline 
{\bf CCD} &V(R)I&&& \struut \\
\hline
& Dec 91 & OBL/FRO & OHP & 7 \struutup\\
& Mar 92 & OBL/LAM  & OHP & 4 \\
& Aug 92 & E. OBLAK & OHP & 6\\
& May 93 & OBL/ZOL & OHP & 6\\
& May 93 & OBL/EMA & LPL & 7\\
& Jun 93 & SEG/LAM  & CAL & 5\\
& Apr 94 & OBL/CMA & OHP & 11\\
& Oct 94 & OBL/CMA & OHP & 11\\
& May 95 & OBL/CMA & OHP & 6\\
& Dec 95 & CHARETON & OHP & 5\\
& Jan 96 & OBL/MOR  & OHP & 13\\
& Nov 96 & OBL/FAL  & OHP & 10\\
& Feb 97 & OBL/CHA  & OHP & 11\\
& Oct 97 & E. OBLAK  & OHP & 11 \\
& Jun 98 & E. OBLAK  & OHP & 9 \struutdown\\
\hline
{\bf CVT} &UBV${\rm B}_{1}{\rm B}_{2}{\rm V}_{1}$G&&& \struut \\
\hline
& Feb 90 & J. CUYPERS  & JFJ & 10 \struutup \\
& Apr 91 & P. LAMPENS  & JFJ & 15 \\
& Jan 92 & P. LAMPENS  & JFJ & 10 \\
& Feb 93 & LAM/HAL  & JFJ & 10\\
& Feb 94 & P. LAMPENS  & JFJ & 10 \\
& Jan 95 & LAM/RUY  & JFJ & 10 \\
& Mar 95 & LAM/RUY  & JFJ & 10 \\
& Feb 97 & P. LAMPENS  & JFJ & 14 \struutdown \\
\hline
\end{tabular} 
\end{table}
\begin{table}[ht]
\caption{{\bf Abbreviation codes for observers}} 
\begin{tabular}{|c|c| }
\hline
\mcol{1}{|c|}{{\bf Code}}                            
& \mcol{1}{c|}{{\bf Observer} \struut} \\
\hline
 CMA & C. MARTIN \\
 CHA & M. CHARETON \\
 EMA & E. MARTIN \\
 FAL & J.L. FALIN \\
 FRO & M. FROESCHLE \\
 HAL & J.L. HALBWACHS \\
 LAM & P. LAMPENS \\
 MOR & G. MORLEY \\
 OBL & E. OBLAK \\
 RUY & G. RUYMAEKERS \\
 SAL & M. SALAMAN \\
 SEG & W. SEGGEWISS \\
 ZOL & E. ZOLA \struutdown \\
\hline
\end{tabular} 
\end{table}


\begin{figure} 
 \resizebox{\hsize}{!}{\includegraphics[width=8.8cm,angle=270]{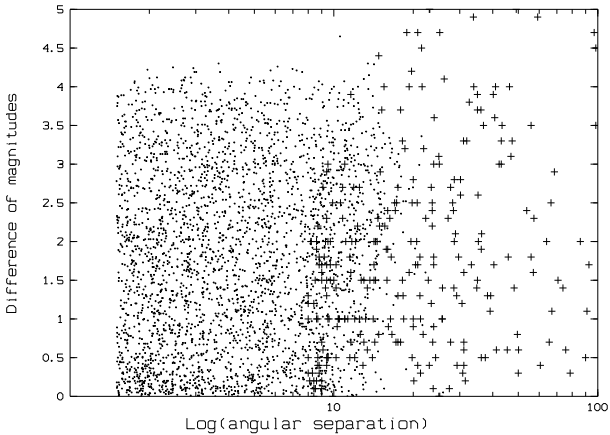}}
 \caption{Sample distribution in angular separation and differential magnitude (dots)
 versus double-star distribution with existing two-component photometry (crosses)} 
 \label{Intro}
\end{figure} 

 It is relevant to recall here the Hipparcos observational strategy in the case of
 adjacent stars (Turon et al \cite{tur92}). With respect to angular separations, systems with separations $< 10$\arcsec~ represent one entry
 only, with the 35\arcsec~ wide IFOV pointing at either the primary, the photocentre or the
 geometric centre, depending on separation and $\Delta {\rm Hp}$. Systems with (maximum) separations $\geq 10$\arcsec~ have
 two or more entries in the Input Catalogue (e.g. a two-pointing double). In such cases an alterning observing 
 strategy has often been used, again depending on $\Delta {\rm Hp}$. Some well-separated components had to be included
 for the purpose of correction only. With respect to magnitudes, the bulk of the stars are brighter than Hp = 10 mag,
 with an upper limit Hp = 12.4 mag (corresponding to V magnitude equal to 12.1 or 12.5 mag depending on the star's colours
 (Grenon et al \cite{gre92}). On the other hand, the Survey is essentially complete within the following magnitude
 limits:\\
 V $\leq 7.9 + 1.1 {\it sin} |{\rm b}|$ for spectral types earlier than or equal to G5,\\
 V $\leq 7.3 + 1.1 {\it sin} |{\rm b}|$ for spectral types later than G5.\\
 Some results of this first astrometric space mission with respect to double stars deserve
 to be mentioned: from the systematic monitoring of a sample of 118.000 stars over 3 years, 
 3000 newly resolved doubles and several thousands of suspected doubles have been detected. In the ($\rho$,
 $\Delta {\rm m}$) plane, the distribution of these new discoveries shows a high concentration in the practically 
 unexplored regime ($\rho < 1$\arcsec, $\Delta {\rm Hp} < 4$ mag)(Fig. 1 in Lindegren et al., \cite{lin97a}). 
 It is also the first time that such a vast material of differential magnitudes (in the Hp passband)
 with a precision of ~0.1 mag has been obtained for double stars of close and intermediate separation.
  
 In terms of physical parameters, what kinds of double stars are actually considered by our programme?
 Limitations on the V magnitude of the primary components are obviously set by the abovecited
 mission constraints. Such limitations imply that, with respect to main sequence primaries,\\
 - there are no faint M dwarfs (with absolute magnitudes $\geq$ 16 mag) in the sample,\\
 - solar-type analogues (absolute magnitudes $\approx$ 5 mag) are included up to some 200 pc,\\
 - dwarfs with spectral types earlier than F5 (absolute magnitudes $\leq$ 4 mag) are included up to 500 pc. \\
 Limitations on the V magnitudes of the secondary component, on the other hand, are governed by the
 observational restriction $\Delta {\rm m} < 3$ mag. This limits the range of the observable mass ratios. 
 Depending on the spectral type, observable values range from unity up to a factor of 2 to 3. \\
 By performing all-sky photometry with the CCD technique, we are entering a regime in angular separation where 
 individual components are now photometrically measured with the same accuracy as with the conventional (photoelectric) 
 technique currently used for joint systems and for individual components of wide pairs. The working area in the
 ($\rho$, $\Delta {\rm m}$) plane is illustrated by Fig.~\ref{Intro}. 
 At the distance of 25 pc, conventionally adopted for the nearby stars, angular separations between 1\arcsec~ and 15\arcsec~ 
 represent true separations ranging from 25 to 375 A.U. Taking an upper distance-limit of 500 pc, these values represent 
 separations beyond 500 A.U. Therefore - though a wide range is covered- our sample specifically addresses
 that part of the distribution in true separation that is longward of the peak value near to 50 A.U.
 (cf. Fig. 1 in Dommanget and Lampens \cite{dom93}). It will furthermore be adequate to investigate the natural 
 "drop-off" for separations larger than some 2000-3000 A.U.\\

\subsection{Summary of campaigns}\label{ssec:sum}

 Observations by this group have been performed in various observatories situated
 in both hemispheres. In the North we observed at Calar Alto (CAL, Spain), 
 Jungfraujoch (JFJ, Switzerland), Observatoire de Haute-Provence (OHP, France), 
 La Palma (LPL, Canary Islands). In the South the La Silla
 observatory was our unique facility (ESO, Chile). A three year lasting
 ESO Key Programme was dedicated to this project for the years 1992-1995 
 (7-009-49K: Periods 49 to 54). Tables~1 and 2 summarize all the campaigns and the used 
 instrumentation for South and North respectively. N.N. is the number of nights. 
 Abbreviation codes for observers can be found in Table~3.\\
 The status of the overall project is presented in Table~4. For these
 statistics, usage was also made of the first large-scale results for
 CCD astrometry and photometry of double stars of the Hipparcos Input
 Catalogue during 1986 and 1987 (Argue et al.~\cite{arg92}; referred to as "the"
 La Palma observations (LPA)). The CCD part of our programme has been completed to 56\% of 
 our initial objective, with a global contribution of 26\% coming from LPA data, another 34\%
 coming from observations by this group and 4\% of data in common.\\
 In the North, the main contribution of 35\% comes from LPA observations while the ESO/OHP/CLA/LPL 
 observations represent 25\% and common stars from both sites contribute another 5\%. The part 
 of not observed northern systems represents another 35\%.\\
 In the South, the ESO observations represent a majority (35\%) while 14\% comes
 from LPA observations and common stars contribute another 4\%. The part of not
 observed southern systems represents 47\%.\\
 Also to be found in Table~4 is the number of observations of multiple systems:
 coverage was achieved for 41\% in the South and only 16\% in the North. At the request 
 of the Hipparcos Double Star Working Group in 1993 an additional 
 set of 81 triple systems with maximum separations between 1\arcsec~ and 15\arcsec,
 but minimum separations (almost) always {\it below 1\arcsec} and no restriction in $\Delta {\rm m}$ 
 (called 'Multiple WG' in Table~4) was observed with a partial coverage of only 31\% . 
 Another one hundred double stars of the Catalogue of Nearby Stars (Jahrei\ss~and Gliese, \cite{jah91}) that were
 not in the Annex of Double and Multiple Stars, have been introduced in the programme lateron, with results to be 
 presented on a independent basis. The overall percentage of absolute photometry for all our southern runs (ESO only) 
 spanning the period Oct.~91 - Jan.~95 is about 40\% of the truly observed time. \\
 Moreover, also the conventional photometric part of our programme at ESO
 suffered heavily from poor climatological conditions: this part has been carried out to 
 as little as 24\% of our initial goal.\\ 

 Some programme stars have been deliberately observed twice while some just happened
 to be in common between our observations and "the" La Palma (LPA) ones. For the CCD programme,
 this amounts to 297 cases out of 1698 for the doubles and to 16 cases out of 162 for
 the multiples. These data are very useful in the assessment 
 of the computed errors for both the astrometry and the photometry.  

\begin{table}[h]
\caption{\bf Status of CCD Observations} 
\setlength{\tabcolsep}{1.5mm}
\vspace{5mm}
\begin{tabular}{|c|cccc| }
\hline 
\mcol{1}{|c|}{{\bf Type}/Hem.}
& \mcol{1}{c}{{\bf Content}}
& \mcol{1}{c}{{\bf Selected}} 
& \mcol{1}{c}{{\bf Observed}}
& \mcol{1}{c|}{ $\%$ \struut}\\
\hline 
{\bf CCD} \struut && Nr & Nr &\\
\hline
  All  & Multiple & 618 & 161  & 26 \struutup \\
       & Double$^{a}$ & 3014  & 1013  & 30+4 \\
       & La Palma$^{a}$ & 3014  &  786 & 22+4 \\
       & common & 3014  &  109 & 4 \\
       & all Doubles & 3014 & 1690  & 56 \struutdown\\
\hline
 Overlap$^{b}$ & Multiple & 81 & 23  & 28 \struutup \\
       & Double$^{a}$ &  375  & 289  & 77 \\
\hline
 North & Multiple & 405 & 65 & 16 \struut \\
       & Double$^{a}$ & 1481 & 441 & 25+5 \\
       & La Palma$^{a}$  & 1481 & 589 & 35+5 \\
       & common  & 1481 &  71 & 5 \\
       & all Doubles & 1481 &  959  & 65 \struutdown \\
\hline
 South & Multiple & 295 & 120  & 41 \struutup \\
       & Multiple WG &  81 &  25  & 31 \struutdown \\
       & Double$^{a}$  & 1908  & 746 & 35+4 \\
       & La Palma$^{a}$  & 1908  & 355 & 14+4 \\
       & common  & 1908  &  81 & 4 \\
       & all Doubles & 1908 & 1020  & 53 \struutdown\\
\hline
\end{tabular} 
$^{a}${\small with the double stars common to our and the LPA campaigns}\\
$^{b}${\small the zones overlap between $\delta_{A} > -10 \degr$ and $\delta_{A} \leq +10 \degr$ }
\end{table}
 

\section{The observational method}\label{sec:met}

\subsection{The common protocol}\label{ssec:pro}
   
   Two different techniques were employed, each of which required 
   a specific protocol to be taken into account by all observers.
   Multi-colour observations have been obtained with the V (occasionally R) and I 
   passbands of the Cousins system or sufficiently close to it. At ESO, CCD
   observations at the Dutch telescope were gathered through a Bessel V and a Gunn
   {\it i} filter.\\
   
   The conventional photometric protocol is a classical one: programme objects 
   and standard stars of the Cousins system taken from a list compiled by M. Grenon (\cite{gre91}) 
   on the basis of lists prepared by Menzies et al. (\cite{men89}, \cite{men91}) 
   and Taylor \& Joner (\cite{tay89}) were alternatively 
   observed in all filters under photometric sky conditions only. At ESO, the same V(R)I 
   filters were used. At Jungfraujoch (Switzerland), the Geneva photometric system with filters 
   UBV${\rm B}_{1}{\rm B}_{2}{\rm V}_{1}$G was employed. The standard star measurements 
   were regularly spaced both in time and in colour.
   Special care was taken when measuring the sky contribution for each component
   (specifically if the angular separation was of the same order as the diaphragm size):
   we tried systematically to measure it diametrically opposed with respect to the other
   component and at about the same distance in angular separation. Observations are
   being reduced in a standard way and results on this part of the programme will be
   reported later on (after the presentation of all our CCD results).\\
   
   A strict protocol was set up concerning the use of the CCD technique:\\
   - we systematically avoided binning;\\
   - sky flat-fields in each filter were taken at the beginning and at the end
     of each observing night; bias frames were taken more regularly during the night;\\ 
   - focus sequences were made for each filter at the start of the night and,
     because of the focus problems (continuous shifts not always related to 
     temperature effects), we frequently had to monitor and adjust the focus during the night in 
     all our runs (compared to the short exposure times used for the acquisition of a single frame, 
     the time needed for adjustment was not negligible, especially for the 0.9m Dutch telescope);\\
   - for the double star programme we used whenever possible a 200x200 pixels window on the CCD chip for
     quick readout and fast file transfer on tape; the full chip was used specifically for the
     flat-fields, the multiple star programme and for the calibration (see Sect.~\ref{ssec:cal} below);\\
   - the 16-bits dynamic range was used whenever possible (requiring e.g. a one bit change option 
     at the Dutch telescope);\\
   - neutral density filters (with magnitude reductions of 2.5 or 5 mag) were employed 
     on both programme and standard stars throughout the night if the former objects
     would have required exposure times below one second without them;\\
   - for each object and each filter, a sequence of multiple exposures was defined 
     with exposure times up to 30~s (sometimes 60~s) adapted so as to have maximum efficiency without
     overexposing the primary component. Mean exposure times are of the order of 10~s. 
     The number of exposures in the sequence was evaluated as $10^{-0.4 \Delta m}$, a function of 
     the catalogued magnitude difference $\Delta m_{AB} = m_{V,B}-m_{V,A}$, with a maximum of 17 
     for $\Delta {\rm m} = 3$ mag.\\
   - as for the conventional programme, standard stars from the list (Grenon
     \cite{gre91}) were taken at regular intervals for the extinction calculation and to
     transform the data to the standard V(R)I system. Two possibilities were considered:\\
   \hspace{5mm} a) under photometric conditions: insertion of few standards (about
    two per hour) permits to obtain standard magnitudes and colours and their differences; \\
   \hspace{5mm} b) under non-photometric conditions: no standard stars observations; instrumental 
    differences of magnitudes and colours only were acquired, implying some loss of accuracy. 

\subsection{The astrometric calibration}\label{ssec:cal}
   A by-product of our CCD observations is the relative geometric configuration
   of the objects. But different campaigns mean different CCD's, so varying scales and also 
   varying orientations. In order to determine the orientation of the CCD stars were trailed 
   over the full length of the chip and a number of "wide" double stars of fixed configuration 
   (separation larger than 10\arcsec) from the lists by Brosche \& Sinachopoulos (\cite{br88}, \cite{br89}) 
   called "astrometric standards" were observed to determine the scale at first. Later we realized 
   that much higher accuracy could be obtained by the inclusion of open star cluster observations (Sinachopoulos 
   et al. \cite{sin93}). Finally, homogeneization of the astrometric data was achieved through direct 
   comparison with the Hipparcos results. The transformation from the local coordinates to Hipparcos 
   coordinates was done by minimizing the squares of all positional differences. Stars with large discrepancies 
   (at the 3$\sigma$-level) were not included in the final calculation of the transformation. This
   allowed the determination of the orientation and of the scale at 0.07\degr, resp. at the 0.01\% level 
   (Oblak et al. \cite{obl97}).\\
   In the near future, we plan to provide a revised list of wide double stars of steady
   configuration because they serve well for a first order approximation of the 
   scale and orientation values and because they are more widely dispersed on the sky 
   than the until now available open clusters.

\subsection{Remarks on the Charge Coupled Devices}\label{ssec:ccd}
   A summary is given in Table~5. Mentioned are the observatory and telescope, the date,
   the general features of the CCD's used in the various campaigns such as identification number,
   pixel width, saturation level. Scale values and zero-points for the orientation will be given later. 
   Worth mentioning is the fact that the stellar images obtained at the Dutch telescope always
   showed ellipticity. This is also taken into account in our reductions.

\begin{table}[h]
\caption{\bf Summary of CCD characteristics } 
\vspace{5mm}
\setlength{\tabcolsep}{1.2mm}
\begin{tabular}{|cccc| }
\hline 
\mcol{1}{|c}{{\bf Site}/Date}
& \mcol{1}{c}{CCD Nr.}
& \mcol{1}{c}{Pxl.w. } 
& \mcol{1}{c|}{Satur. \struut}\\
\hline 
{\bf ESO} & DUT 0.9m &($\mu$)&(KADU) \struut \\
\hline
 Oct 91 - Nov 92 & 7  & 22 & 16.8\struutup \\
 Mar 93 & 5  & 30  & 50\\
 Aug 93 - Dec 93 & 29 & 27 & 32 \\
 May 94 & 5  & 30  & 50\\
 Aug 94 & 29 & 27 & 64  \\
 Nov 94 - Jan 95 & 33 & 27 & 64 \struutdown\\
\hline 
 {\bf OHP} & 0.8/1.2m &($\mu$)&(KADU) \struut \\
\hline 
         & RCA1 & 30 & 22 \struutup \\
 91-98 & RCA2 & 30 & 19  \\
         & TK512 & 27 & 32  \\
\hline 
 {\bf CLA} & 1.2m & ($\mu$)&(KADU) \struut \\
\hline 
 93 & TEK4 & 27 & 18  \\
\hline 
 {\bf LPA} & 1.0m & ($\mu$)&(KADU) \struut \\
\hline 
 93 & EEV7 & 22 & 64  \\
\hline
\end{tabular} 
\end{table}


\section{The reduction method}\label{sec:red}
\subsection{Pre-reduction}
  The raw CCD images in the various filters are treated in a standard way, 
  i.e. bias subtracted and flat-fielded. Since our programme stars are bright,
  the results are not very sensitive to the choice of the used flat-fields. Median
  (sky) flats normalized to unit intensity over a couple of nights within one mission
  serve their purpose well as long as care is taken that identical observing conditions
  prevailed (no insertion/removal of neutral density filters; no CCD dismounting;
  no filter cleaning operation, etc.)
 
\subsection{Reduction of differential measurements}\label{ssec:diff}
  Various packages exist for the accurate astrometric and photometric reduction of CCD 
  images in crowded fields: they take advantage of some well-exposed, isolated stars
  in the field of interest. The larger the number of isolated stars, the better
  the accuracies (e.g. DAOPHOT developed and distributed by P. Stetson (\cite{ste87})).
  However they cannot be applied here: the limited chip size coupled to the brightness
  of the objects makes that the majority of our frames show two overlapping
  profiles, usually without any other star in the field. A direct profile 
  fitting method, that allows the separation of the individual
  profiles of the closest pairs, i.e. the pairs with separations somewhat larger
  than the width of the seeing disk, is thus desirable.\\
  A one dimensional profile fitting method has been used by Sinachopoulos
  (\cite{sin92}): it implies fitting a Franz profile to a row and a column projection for each frame
  via a least squares method supported by an expert system. It was applied to double stars
  having separations generally larger than 5\arcsec.
  The reduction of closer pairs observed at La Palma was done
  by applying another method based on centroids and isophotes (Irwin \cite{irw85}).
  Our approach is different: a specific two-dimensional profile fitting method
  was developed within the MIDAS software package (Cuypers \cite{cuy97}). This reduction method 
  fits with a least squares technique a bidimensional Moffat profile (Moffat \cite{mo82}) with elliptical isophotes 
  to all the components simultaneously. Since the angular separation of the components 
  is generally less than the size of the isoplanatic area, the shape of the point spread function
  can indeed be considered identical for all components. This method favourably compared to DAOPHOT 
  and was succesfully applied to systems with up to 10 "components" (Lampens \& Seggewiss \cite{lam95}). 
  The data obtained consists, after sky subtraction, of relative positions and differential 
  magnitudes in each of the filters. 

\subsection{Reduction of magnitudes in a standard photometric system}\label{ssec:stan}
  
  We already reported that during nights of good photometric quality
  standard stars in the Cousins system have been observed
  (see Sect.~\ref{sec:met}). These stars were used in a least squares model to
  derive extinction values for each night and transformation
  coefficients from the instrumental to the standard system. The
  differences between both pairs of standard passbands (V,{\it i}){\em
  (Bessell/Gunn)} vs.\ (V,I){\em (Cousins)} are not negligible but
  they are safely treated by the transformation equations.\\

	 
  If possible, several nights of the same observing campaign were
  reduced simultaneously in order to obtain more accurate
  values for the zeropoints and transformation coefficients. Nights with 
  neutral density filters were always treated separately.
  Colour corrections were linear. A breakpoint in colour was introduced when necessary
  in order to have the possibility of using different colour corrections for blue and
  red stars.\\

  In general, transformation errors were shown to be of the order of 0.02-0.03 mag (Lampens
  et al. \cite{lam197}).\\

  
  Instrumental global magnitudes (for the system) have also been computed for all observed programme
  stars. Individual component magnitudes were then derived from these and the differential magnitudes
  obtained earlier. All component magnitudes have been subsequently transformed into the Cousins standard 
  system, after correction for the extinction. Errors on the CCD global magnitudes are comparable
  to those from photoelectric photometry (a few millimag in good conditions): they come from the photon statistics.
  On the other hand, errors on the differential magnitudes are introduced through the fitting procedure and are 
  somewhat larger: they are much reduced by taking a series of exposures (and depend
  on angular separation as well as on the difference itself, Lampens et al. \cite{lam197}). These two error 
  sources contribute differently to the errors on the component magnitudes that will be listed in the 
  forthcoming papers of the series.
  
 

\section{Prospects and conclusions}\label{sec:con}
  
\subsection{First results and prospects}\label{ssec:res}

   First but preliminary results have already been presented on various occasions (Oblak et al. \cite{obl293}; 
    Oblak et al. \cite{obl96}; Lampens et al. \cite{lam197}; Oblak et al. \cite{obl97}). 
   The data papers should soon follow. A detailed comparison with the recently published Hipparcos
   data will be done after completion of the reduction for all missions. In addition, we will publish a list 
   of astrometric calibration double stars: these are the wide double stars of our lists with steady configuration 
   that can be used anywhere on the sky for estimating the scale and the orientation of CCD cameras.

\subsection{General conclusions}\label{ssec:gen}
\begin{enumerate}

\item We obtained high-quality astrometry and all-sky multi-colour photometry for the components of intermediate visual 
  double stars using small telescopes of the 1m class equipped with a CCD. The final accuracy level 
  of our data matches that of the Hipparcos mission for systems with angular separations in the range 
  2\arcsec~ to 12\arcsec. This has been possible thanks to the introduction and application of a strict protocol 
  combined with the common usage of a dedicated reduction procedure.\\
\item   We used this specific reduction tool for images containing up to 10 "components" and for
  separations ranging from very large (over 30\arcsec) to very close, i.e. down to the limit imposed by 
  the seeing disk. The method proved to be superior to a reduction by means of the cluster
  reduction package DAOPHOT (also available in the MIDAS software).\\
\item   We were able to provide a priori ground-based values for the geometric 
  configuration of hundreds of "intermediate" visual double and multiple systems
  for the HIPPARCOS double star reduction: for several systems with angular separations 
  above 1\arcsec, this new information, given along with the individual colour indices 
  of the components, allowed to remove the ambiguities generated by the grid step 
  ($\sim$ 1.2\arcsec) and thus to improve the reliability of the Hipparcos Catalogue solution.
  These results were also valuable for providing a good starting point during the 
  double star re-reduction by the Hipparcos Reduction Consortia (Falin \& Mignard \cite{fal98}).\\ 
\item   Such observations allow to deduce a very sparsely known quantity among close visual binaries:
  the colour difference between individual components. This information has been very useful
  for comparing with the data from the Hipparcos mission. Component colour indices are important 
  astrophysical parameters that are missing in too many researches on double and multiple stars: however 
  they are easily and accurately obtained for those systems with separations in the intermediate range with 
  simple means in excellent photometric conditions. The determination of the distribution 
  functions of true separations, mass and luminosity ratios, differences of temperatures 
  for a significant sample of double stars are obvious future applications, profitable in
  the domains of e.g. stellar formation and modeling of double stars. Of course, we will include the
  $\approx$ 11\% of (wide) systems already having good and complete photometry in such studies.

\end{enumerate}      

\begin{acknowledgements}
  We gratefully acknowledge the allocation of telescope time by the European Southern Observatory as well as
  the Observatoire de Haute-Provence and the Calar Alto, La Palma and Jungfraujoch Observatories
  during the full length of the programme. The network {\it R\'eseau Europ\'een des Laboratoires:
  Etoiles Doubles Visuelles} was supported in 1992 by the French {\it Minist\`ere de la Recherche 
  et de la Technologie}. We particularly thank J.L. Falin of the FAST reduction team for communicating some
  preliminary Hipparcos results. We appreciate the help of our colleagues N. Argue, P. Brosche, J. Dommanget,
  A. Duquennoy, G. Jasniewicz, M. Geffert, M. Grenon, J.C. Mermilliod and F. Mignard for helpful discussions within the Network.
  We thank the referee, Prof. P. Brosche, for many helpful suggestions. EO acknowledges financial
  support from the French {\it Minist\`ere des Affaires Etrang\`eres} for the programmes {\it Alliance} with the United 
  Kingdom and {\it Procope} with Germany. PL and JC acknowledge funding by project G.0265.97 of the {\it Fonds voor Wetenschappelijk
  Onderzoek} (FWO).  
\end{acknowledgements}

\end{document}